\documentclass[12pt]{article}
\usepackage{sc3conf}
\begin{document}

\def\cstok#1{\leavevmode\thinspace\hbox{\vrule\vtop{\vbox{\hrule\kern1pt
\hbox{\vphantom{\tt/}\thinspace{\tt#1}\thinspace}}
\kern1pt\hrule}\vrule}\thinspace}

\title{New Photon Propagators in Quantum Electrodynamics}

\authors{G. Esposito,\adref{1}}

\addresses{\1ad Istituto Nazionale di Fisica Nucleare, Sezione di Napoli,
Complesso Universitario di Monte S. Angelo, Via Cintia,
Edificio N', 80126 Napoli, Italy.}

\maketitle

\begin{abstract} 
A Lagrangian for quantum electrodynamics is
found which makes it explicit that the photon mass is eventually
set to zero in the physical part on observational ground.
It remains possible to obtain a counterterm Lagrangian where the
only non-gauge-invariant term is proportional to the squared
divergence of the potential, while the photon propagator in
momentum space falls off like $k^{-2}$ at large $k$, which indeed
agrees with perturbative renormalizability. The resulting radiative
corrections to the Coulomb potential in QED are also shown to be
gauge-independent. A fundamental role of the space of
$4$-vectors with components given by $4 \times 4$ matrices is
therefore suggested by our scheme, where such matrices can be used 
to define a single gauge-averaging functional in the path integral.
\end{abstract}

\section{Introduction}

Our research on quantum electrodynamics (hereafter QED) has
been motivated by the need to understand how to quantize gauge
theories of fundamental interactions in case the Higgs boson
were to remain elusive. Although we have failed in this respect,
we have found a number of field-theoretical properties which
seem to be of interest and are now described, relying upon
Ref. [1].

To begin, we recall that gauge-invariant Lagrangians are
naturally considered to generate mass via the Higgs mechanism,
if fundamental scalar fields are taken to exist. 
But gauge-invariant Lagrangians suffer from a ``degeneracy''
in that they lead to non-invertible operators on gauge fields. At
about the same time when Higgs was elaborating his model, it became
clear thanks to Feynman, DeWitt, Faddeev and Popov (and various
other authors after them) that the appropriate Lagrangian contains
actually (at least) 3 ingredients:
\vskip 0.3cm
\noindent
(i) A gauge-invariant part ${\cal L}_{\rm inv}$;
\vskip 0.3cm
\noindent
(ii) A term called {\it gauge-breaking}, {\it gauge-fixing}
or {\it gauge-averaging};
\vskip 0.3cm
\noindent
(iii) Contribution of ghost fields [they have classical roots;
just think of the need to preserve the supplementary condition
under gauge transformations, which leads to a gauge function
ruled by a differential operator].
\vskip 0.3cm
\noindent
The resulting physical predictions are independent of the particular
supplementary condition and of any choice for the matrix of
gauge parameters.

Thus, a gauge-invariant Lagrangian ${\cal L}_{\rm inv}$ with scalars 
leads to the Higgs mechanism for mass generation. But if no
fundamental scalar field exists, we face the problem of studying
a Becchi--Rouet--Stora--Tyutin-invariant Lagrangian and understanding
under which conditions this is compatible with mass terms
(which is not the same as generating mass!).
 
\section{New gauges for Maxwell and QED}

Our first basic remark is that
\begin{equation}
A_{\mu}A^{\mu}=g^{\mu \nu}A_{\mu} A_{\nu}
={1\over 4}{\rm Tr}(\gamma^{\mu}\gamma^{\nu})A_{\mu}A_{\nu}.
\end{equation}
We have therefore looked, in a first moment, for a linear gauge 
combining the effect of Lorenz gauge [2]
and $\gamma$-matrices. 
However, one cannot simply add the derivatives
of $A_{\mu}$ in the Lorenz gauge and $\gamma^{\mu}$ terms, since
the latter are $4$-vectors with components given by $4 \times 4$
matrices. The only well defined operation on such objects is
the one giving rise to the matrix
\begin{equation}
\Phi_{i}^{\; j}(A) \equiv 
\Bigr(\delta_{i}^{\; j}\partial^{\mu}
+\beta (\gamma^{\mu})_{i}^{\; j}\Bigr)A_{\mu}(x),
\end{equation}
where the parameter $\beta$ is introduced to ensure that all
terms in $\Phi_{i}^{j}$ have the same dimension (i.e.
$\beta$ has dimension [length]$^{-1}$). 
There is only one coefficient, $\beta$, since
only one potential $A_{\mu}$ is available for contraction with
$\gamma^{\mu}$ in the Abelian case. 

The resulting gauge-averaging term in the Lagrangian is taken to be
\begin{equation}
{\cal L}_{G.A.}={\Phi^{2}(A)\over 2\alpha}
={1\over 2\alpha}\Phi_{i}^{\; j}(A) \Omega_{j}^{\; k}
\Phi_{k}^{\; i}(A),
\end{equation}
where we have defined the symmetric matrix
\begin{equation}
\Omega_{j}^{\; k} \equiv {1\over 4} 
\delta_{j}^{\; k} .
\end{equation}
Therefore ($g^{\mu \nu}=-{1\over 4}{\rm Tr}(\gamma^{\mu}\gamma^{\nu})$
and $\beta \rightarrow \pm i \beta$ in Euclidean theory)
\begin{eqnarray}
\; & \; & \Phi_{i}^{\; j}(A) \Omega_{j}^{\; k}
\Phi_{k}^{\; i}(A)
={1\over 4} \Bigr[4(\partial^{\mu}A_{\mu})(\partial^{\nu}A_{\nu})
+\beta (\partial^{\mu}A_{\mu})(\gamma^{\nu})_{i}^{\; i}A_{\nu}
+\beta (\gamma^{\mu})_{i}^{\; i}A_{\mu}\partial^{\nu}A_{\nu} 
\nonumber \\
&+& \beta^{2}(\gamma^{\mu})_{i}^{\; j}
(\gamma^{\nu})_{j}^{\; i}
A_{\mu}A_{\nu}\Bigr]
=(\partial^{\mu}A_{\mu})(\partial^{\nu}A_{\nu})
+\beta^{2} A_{\mu}A^{\mu}.
\end{eqnarray}
It should be stressed that the matrix $\Phi_{i}^{\; j}(A)$ is
a tool to express in a concise and elegant form
the gauge-averaging term in the full action, but 
{\it our gauge-averaging functional for} QED {\it is not a matrix}
and is equal to
\begin{equation}
\Phi(A) \equiv \left(\Phi_{i}^{\; j}(A)\Omega_{j}^{\; k}
\Phi_{k}^{\; i}(A)\right)^{{1\over 2}}.
\end{equation}
One cannot regard $\Phi_{i}^{\; j}(A)$ itself as a gauge-averaging
functional, since otherwise one would get 16 supplementary
conditions which are totally extraneous to the quantum (as well
as classical) theory. One can however say that the matrix (2) acts
as a ``potential'' for the gauge-averaging functional, in that
the definition (6) can be given. The price to be paid is lack of
a simple formula for the ghost operator $\cal P$, whose action
is given by
\begin{equation}
{\cal P}: \varepsilon \rightarrow 
- \left[{(\partial^{\nu}A_{\nu})\over \Phi(A)}
\partial^{\mu}\partial_{\mu}
+{\beta^{2}A^{\mu}\over \Phi(A)}\partial_{\mu}\right]
\varepsilon , 
\end{equation}
which however does not spoil
the occurrence of $A_{\mu}A^{\mu}$ terms in the full action functional.
Thus, the starting point remains the well known property that only
one gauge-fixing condition ($\Phi(A)=\zeta$) is needed, and we choose
$\Phi(A)$ in the non-linear form (6).

\section{Bare photon propagator}

The photon propagator is obtained by first considering the
gauge-field operator $P_{\mu \nu}$ from the path integral,
then taking its symbol $\sigma(P_{\mu \nu}) 
\equiv \sigma_{\mu \nu}$, and eventually inverting 
$\sigma$ to find $\widetilde \sigma$ such that
\begin{equation}
\sigma_{\mu \nu}{\widetilde \sigma}^{\nu \lambda}
=\delta_{\mu}^{\; \lambda}.
\end{equation}
This leads to
\begin{equation}
\bigtriangleup^{\mu \nu}(x,y)= \int_{\zeta} 
{d^{4}k \over (2\pi)^{4}} \;
{\widetilde \sigma}^{\mu \nu}(k)e^{ik \cdot (x-y)}.
\end{equation}
We find, by virtue of our gauge-averaging term,
\begin{equation}
{\cal L}=\partial^{\mu}\rho_{\mu}+{1\over 2}A^{\mu}P_{\mu \nu}A^{\nu},
\end{equation}
where
\begin{equation}
\rho_{\mu} \equiv -{1\over 2}A_{\nu}\partial^{\nu}A_{\mu}
+{1\over 2}A_{\nu}\partial_{\mu}A^{\nu}
+{1\over 2\alpha}A_{\mu}\partial^{\nu}A_{\nu},
\end{equation}
\begin{equation}
P_{\mu \nu} \equiv g_{\mu \nu} \left[-\cstok{\ }+{\beta^{2}\over \alpha}
\right]+\left(1-{1\over \alpha}\right)\partial_{\mu}\partial_{\nu}.
\end{equation}
Of course, ${\rm div}\rho$ does not affect the field equations,
and $P_{\mu \nu}$ depends on gauge parameters.
In particular, we here choose for simplicity:
\begin{equation}
P_{\mu \nu}(\alpha=1)=g_{\mu \nu}\left(-\cstok{\ }+\beta^{2}\right),
\end{equation}
which implies
\begin{equation}
\sigma(P_{\mu \nu}(\alpha=1))=(k^{2}+\beta^{2})g_{\mu \nu},
\end{equation}
with Euclidean photon propagator
\begin{equation}
\bigtriangleup_{E}^{\mu \nu}(x,y)= \int_{\Gamma}
{d^{4}k \over (2\pi)^{4}}
{g^{\mu \nu}\over (k^{2}+\beta^{2})}e^{ik \cdot (x-y)}.
\end{equation}
Integration along the real axis avoids poles of the integrand,
and the interpretation of the massive term will become clear
in the following sections.

\section{Perturbative renormalization}

According to the perturbative renormalization programme, we now
distinguish between bare quantities, denoted by the $B$ subscript,
and physical quantities (which do not carry any subscript).
This is done for all fields, physical parameters and gauge 
parameters by assuming that multiplicative renormalizability still
holds with our gauge (6), so that we can exploit the relations
\begin{equation}
(A_{\mu})_{B}=\sqrt{z_{A}}\; A_{\mu}, \; \Longrightarrow 
F_{B}^{\mu \nu}=\partial^{\mu}A_{B}^{\nu}
-\partial^{\nu}A_{B}^{\mu},
\end{equation}
\begin{equation}
\psi_{B}=\sqrt{z_{\psi}} \; \psi,
\end{equation}
\begin{equation}
m_{B}={z_{m}\over z_{\psi}}m,
\end{equation}
\begin{equation}
e_{B}={z_{e}\over z_{\psi}\sqrt{z_{A}}}e,
\end{equation}
\begin{equation}
\alpha_{B}={z_{A}\over z_{\alpha}}\alpha,
\end{equation}
\begin{equation}
\beta_{B}=\rho \beta,
\end{equation}
where the $\rho$ coefficient will be fixed shortly.
Hence we find
\begin{equation}
{\cal L}-{\cal L}_{\rm gh}
={\cal L}_{\rm {ph}}+{\cal L}_{\rm {ct}},
\end{equation}
where the physical part reads 
\begin{eqnarray}
{\cal L}_{\rm ph}&=&-{1\over 4}F_{\mu \nu}F^{\mu \nu}
+{\overline \psi}i\gamma^{\mu}\partial_{\mu}\psi
-e{\overline \psi}\gamma^{\mu}A_{\mu}\psi
-m{\overline \psi}\psi \nonumber \\
&-&{1\over 2\alpha}(\partial^{\mu}A_{\mu})^{2}
-{\beta^{2}\over 2\alpha}A_{\mu}A^{\mu},
\end{eqnarray}
with
\begin{equation}
{\beta^{2}\over \alpha} \equiv m_{\gamma}^{2},
\end{equation}
while the part involving counterterms is given by 
\begin{eqnarray}
{\cal L}_{\rm ct}&=& -{1\over 4}(z_{A}-1)F_{\mu \nu}F^{\mu \nu}
+(z_{\psi}-1){\overline \psi}i \gamma^{\mu}\partial_{\mu}\psi
\nonumber \\
&-& (z_{e}-1)e {\overline \psi}\gamma^{\mu}A_{\mu}\psi 
-(z_{m}-1)m{\overline \psi}\psi \nonumber \\
&-& {1\over 2\alpha}(z_{\alpha}-1)(\partial^{\mu}A_{\mu})^{2}
-{\beta^{2}\over 2 \alpha}(\rho^{2}z_{\alpha}-1)A_{\mu}A^{\mu}. 
\end{eqnarray}
Note that, if
\begin{equation}
\rho={1\over \sqrt{z_{\alpha}}},
\end{equation}
the counterterm Lagrangian reduces to the familiar form
in the Lorenz gauge, and the renormalization of $\beta$
is not independent of the renormalization of $\alpha$,
in agreement with (24). At this stage we deal with a freely
specifiable gauge parameter, i.e. $\alpha$, and with the
physical mass parameter $m_{\gamma}$.

\section{Radiative corrections in QED}

Since
\begin{equation}
{\beta_{B}^{2}\over \alpha_{B}}={\rho^{2}\beta^{2}\over
{z_{A}\over z_{\alpha}}\alpha}={m_{\gamma}^{2}\over z_{A}}
\equiv {\widetilde m}_{\gamma}^{2},
\end{equation}
we find in the bare theory
\begin{equation}
\sigma_{\mu \nu}(k)=\left(g_{\mu \nu}-{k_{\mu}k_{\nu}\over k^{2}}
\right)(k^{2}+{\widetilde m}_{\gamma}^{2})
+{k_{\mu}k_{\nu}\over k^{2}}{1\over \alpha_{B}}
(k^{2}+\alpha_{B} {\widetilde m}_{\gamma}^{2}),
\end{equation}
with inverse
\begin{equation}
{\widetilde \sigma}^{\mu \nu}(k)={g^{\mu \nu}\over
(k^{2}+{\widetilde m}_{\gamma}^{2})}
+{(\alpha_{B}-1)k^{\mu}k^{\nu}\over 
(k^{2}+\alpha_{B} {\widetilde m}_{\gamma}^{2})
(k^{2}+{\widetilde m}_{\gamma}^{2})}.
\end{equation}
Note that ${\widetilde \sigma}^{\mu \nu}$ falls off like $k^{-2}$
at large $k$, in agreement with perturbative renormalizability.

In the renormalized theory one has on general ground
\begin{equation}
\Sigma_{\mu \nu}(k)=g_{\mu \nu}u_{1}(k^{2})+k_{\mu}k_{\nu}
u_{2}(k^{2}),
\end{equation}
\begin{equation}
{\widetilde \Sigma}^{\mu \nu}(k)=g^{\mu \nu}d_{1}(k^{2})
+k^{\mu}k^{\nu}d_{2}(k^{2}),
\end{equation}
while diagrammatic analysis shows that
\begin{equation}
{\widetilde \Sigma}^{\mu \nu}(k)={\widetilde \sigma}^{\mu \nu}(k)
+{\widetilde \sigma}^{\mu \lambda}\Pi_{\lambda \rho}
{\widetilde \Sigma}^{\rho \nu}(k),
\end{equation}
with $\Pi_{\lambda \rho}$ the polarization tensor, given by
\begin{eqnarray}
\; & \; & \Pi_{\mu \nu}(k)=\sigma_{\mu \nu}(k)-\Sigma_{\mu \nu}(k) 
\nonumber \\
&=& g_{\mu \nu}\left(k^{2}+{\widetilde m}_{\gamma}^{2}
-u_{1}\right)+k_{\mu}k_{\nu}\left({1\over \alpha_{B}}-1-u_{2}\right) 
\nonumber \\
&=& g_{\mu \nu}a_{1}(k^{2})+k_{\mu}k_{\nu}a_{2}(k^{2}).
\end{eqnarray}
Current conservation implies that
$k^{\mu}\Pi_{\mu \nu}=0$, which leads to
\begin{equation}
a_{1}=-k^{2}a_{2},
\end{equation}
\begin{equation}
u_{1}+k^{2}u_{2}={1\over \alpha_{B}}
(k^{2}+\alpha_{B} {\widetilde m}_{\gamma}^{2}),
\end{equation}
and hence
\begin{equation}
\Pi_{\mu \nu}(k)=\left(g_{\mu \nu}-{k_{\mu}k_{\nu}\over k^{2}}
\right)\left(k^{2}+{\widetilde m}_{\gamma}^{2}-u_{1}\right).
\end{equation}
This is in gauge-independent form, as expected.

For example, we may set
\begin{equation}
u_{1}={\widetilde m}_{\gamma}^{2}+f(k^{2}),
\end{equation}
which implies
\begin{equation}
u_{2}={1\over \alpha_{B}}-{f(k^{2})\over k^{2}}.
\end{equation}
Therefore
\begin{equation}
\Sigma_{\mu \nu}(k)=\left(g_{\mu \nu}-{k_{\mu}k_{\nu}\over k^{2}}
\right)\left(f(k^{2})+{\widetilde m}_{\gamma}^{2} \right)
+{k_{\mu}k_{\nu}\over k^{2}}{1\over \alpha_{B}}
(k^{2}+ \alpha_{B} {\widetilde m}_{\gamma}^{2}).
\end{equation}
Thus, the coefficient of the longitudinal part 
${k_{\mu}k_{\nu}\over k^{2}}$ is the same in the bare as well as
in the full theory, while the coefficients of the transverse
part $g_{\mu \nu}-{k_{\mu}k_{\nu}\over k^{2}}$ depend on $\alpha_{B}$
and $\beta_{B}$ in such a way that the difference $\sigma_{\mu \nu}
-\Sigma_{\mu \nu}$ is indeed gauge-independent:
\begin{equation}
\Pi_{\mu \nu}(k)
=\left(g_{\mu \nu}-{k_{\mu}k_{\nu}\over k^{2}}\right)
(k^{2}-f(k^{2})).
\end{equation}
The renormalized photon propagator in momentum space is the 
inverse of $\Sigma_{\mu \nu}(k)$, i.e.
\begin{equation}
{\widetilde \Sigma}^{\mu \nu}(k)={g^{\mu \nu}\over
\left(f(k^{2})+{\widetilde m}_{\gamma}^{2}\right)}
+{\left(\alpha_{B} {f(k^{2})\over k^{2}}-1 \right)k^{\mu}k^{\nu}
\over (k^{2}+\alpha_{B} {\widetilde m}_{\gamma}^{2})
\left(f(k^{2})+{\widetilde m}_{\gamma}^{2}\right)}.
\end{equation}

As an application, we consider radiative corrections to Coulomb's law:
\begin{equation}
{\cal A}^{0}=A^{0}+{\widetilde \Sigma}^{0 \rho}
\Pi_{\rho \lambda}A^{\lambda},
\end{equation}
where 
\begin{equation}
A^{0}={\rm const.}  {q\over k^{2}},
\end{equation}
while
\begin{equation}
{\widetilde \Sigma}^{0 \rho}\Pi_{\rho \lambda}
=\left(\delta_{\; \lambda}^{0}-{k^{0}k_{\lambda}\over k^{2}}
\right){(k^{2}-f(k^{2}))\over \left(f(k^{2})
+{\widetilde m}_{\gamma}^{2}\right)},
\end{equation}
since $k^{0}k^{\rho}d_{2}(k^{2})\Pi_{\rho \lambda}=0$. 
In other words, we find that the renormalized potential 
${\cal A}^{0}$ depends on gauge parameters $\alpha, \beta$ not
separately, which would have led to unavoidable gauge dependence
(since $\beta =m_{\gamma}\sqrt{\alpha}$), but only through the
product ${1\over z_{A}} {\beta^{2}\over \alpha}$. The latter is 
proportional to the photon mass parameter $m_{\gamma}^{2}$ in the physical
Lagrangian of perturbative renormalization. Thus, {\it the resulting
short-range potential only depends on a mass parameter in the
physical Lagrangian and is therefore, with the above understanding,
gauge independent}. Note that the classical long-range part 
${q\over k^{2}}$ resulting from $A^{0}$ is still present, 
and eventually our
${\widetilde m}_{\gamma}$ is set to zero on observational ground. 

Note also that, in the light of previous remarks, the most
general form of $u_{1}$ is
\begin{equation}
u_{1}=u_{1} \left(k^{2};{\beta_{B}^{2}\over \alpha_{B}}\right)
=u_{1}(k^{2};{\widetilde m}_{\gamma}^{2}),
\end{equation}
leading to 
\begin{equation}
d_{1}={1\over u_{1}(k^{2};{\widetilde m}_{\gamma}^{2})},
\end{equation}
and
\begin{equation}
d_{2}={1\over k^{2}}\left[{\alpha_{B} \over 
(k^{2}+\alpha_{B} {\widetilde m}_{\gamma}^{2})}
-{1\over u_{1}(k^{2}; {\widetilde m}_{\gamma}^{2})}\right].
\end{equation}
Once more, only $d_{2}$ is gauge-dependent, since
its first term depends on $\alpha_{B}$. 

\section{Results and open problems}

We have not truly generated mass for the photon, but we have
developed tools for a more thorough treatment of its massless
nature. More precisely, our original results are as follows.
\vskip 0.3cm
\noindent
(i) Derivation of the gauge-averaging functional (6) for QED.
\vskip 0.3cm
\noindent
(ii) New photon propagators in quantum electrodynamics, with 
$m_{\gamma}$ as an explicit mass parameter in the physical Lagrangian.
\vskip 0.3cm
\noindent
(iii) Renormalization of the gauge parameter $\beta$ in such a way that
the counterterm Lagrangian has only one term which is not
gauge-invariant. 
\vskip 0.3cm
\noindent
(iv) Renormalized photon propagator in our gauges,
and proof of gauge independence of the associated short-range potential,
adding evidence in favour of our model being physically relevant.
\vskip 0.3cm

Objections (O) can be raised and answers (A) can be given
along the following lines:
\vskip 0.3cm
\noindent
(O1) Precisely in the Abelian case, the Higgs--Kibble (HK) model
has a mass term which is cohomologically non-trivial and hence
it has physical content.
Such a property is hidden by your model.
\vskip 0.3cm
\noindent
(A1) The HK model relies upon fundamental scalar fields, whereas
we have tried to understand what happens if such scalar fields
do not exist.
\vskip 0.3cm
\noindent
(O2) The approach presented is, eventually, very phenomenological,
in that suitable combinations of gauge parameters are used to fit
the experimental data.
\vskip 0.3cm
\noindent
(A2) Yes indeed, without fundamental scalar fields there remains
a fundamental procedure, i.e. construction of a Lagrangian leading
to invertible operators on the potentials, but then we end up
fixing physical parameters on observational ground. It might be
acceptable to the extent that we are satisfied with
perturbative renormalization.

Note also that in the massive QED model, which is ruled
by the field equations
\begin{equation}
(i\gamma^{\mu}\partial_{\mu}-M I)\psi=e A^{\mu}\gamma_{\mu}\psi,
\end{equation}
\begin{equation}
\partial^{\mu}F_{\mu \nu}+m^{2}A_{\nu}=-e {\overline \psi}
\gamma_{\nu} \psi,
\end{equation}
the photon propagator is given by
\begin{equation}
\bigtriangleup^{\mu \nu}(x,y)=\int {d^{4}k \over (2\pi)^{4}}
\left(g^{\mu \nu}-{k^{\mu}k^{\nu}\over m^{2}}\right)
{-i \over (k^{2}-m^{2}+i \varepsilon)}e^{ik \cdot (x-y)}.
\end{equation}
Its integrand is constant at large $k$, and this leads to
a non-renormalizable theory, in which the divergence of a
Feynman diagram increases with the number of internal photon
lines. Such unpleasant features are not shared 
by our photon propagator (in
agreement with our analysis of perturbative renormalizability),
which keeps the standard $k^{-2}$ behaviour at large $k$.

The main open problem is as follows. On the one hand,
no gauge-averaging is needed in the path integral for fermionic
fields. Thus, by construction, our approach does not generate 
masses for fermions, and hence does not provide an alternative
to the Higgs mechanism in the standard model. On the other hand,
we end up by putting the emphasis on the space of $4$-vectors
with components given by $4 \times 4$ matrices, which is a
natural structure for theories incorporating fermions. 

As far as we can see, further research topics are suggested
by our approach, in particular:
(i) How to prove explicitly perturbative renormalizability 
in the non-Abelian case in our broader framework;
(ii) Possible occurrence of Gribov ambiguities in the
non-perturbative formulation of non-Abelian theories;
(iii) Evaluation of mass terms for ghost fields;
(iv) Equations for Green functions in QED and quantum
Yang--Mills along the lines of Gribov [3]. His idea was
to formulate equations for Green's functions and vertices in
a form that does not contain any divergences, so that solutions
of these equations will be able to account for both perturbative
and non-perturbative phenomena.

\end{document}